\documentclass[aps,twocolumn,prx,floatfix,showpacs,citeautoscript,superscriptaddress,longbibliography]{revtex4-2}

\usepackage{amssymb}
\usepackage{amsmath}
\usepackage{graphicx}
\usepackage[caption=false]{subfig}
\usepackage{url}
\usepackage{xcolor}
\usepackage{dcolumn}
\usepackage[colorlinks, linkcolor=blue, citecolor=blue, urlcolor=black]{hyperref}
\usepackage{siunitx}

\begin{document}

\title{Coherent Optical Control of Electron Dynamics in Patterned Graphene Nanoribbons}
\author{Riek H. Rüstemeier}
\affiliation{Center for Optical Quantum Technologies and Institute for Quantum Physics, University of Hamburg, 22761 Hamburg, Germany}
\author{H. P. Ojeda Collado}
\affiliation{Center for Optical Quantum Technologies and Institute for Quantum Physics, University of Hamburg, 22761 Hamburg, Germany}
\affiliation{The Hamburg Center for Ultrafast Imaging, Luruper Chaussee 149, 22761 Hamburg, Germany}
\date{\today}
\author{Ludwig Mathey}
\affiliation{Center for Optical Quantum Technologies and Institute for Quantum Physics, University of Hamburg, 22761 Hamburg, Germany}
\affiliation{The Hamburg Center for Ultrafast Imaging, Luruper Chaussee 149, 22761 Hamburg, Germany}
\date{\today}

\begin{abstract}
The field of coherent electronics aims to advance electronic functionalities by utilizing quantum coherence.  
Here, we demonstrate a viable and versatile methodology for controlling electron dynamics optically in graphene nanoribbons.  
In particular, we propose to flatten the band structure of armchair graphene nanoribbons via control electrodes, arranged periodically along the extended direction of the nanoribbon. 
This addresses a key mechanism for dephasing in solids, which derives from the momentum dependence of the energy gap between the valence and the conduction band.  
We design an optimal driving field pulse to produce collective Rabi oscillations between these bands, in their flattened configuration.
As an example for coherent control, we show that these optimized pulses can be used to invert the entire electronic band population by a $\pi$ pulse in a reversible fashion, and to create a superposition state via a $\pi/2$ pulse, which generates an alternating photocurrent.
Our proposal consists of a platform and methodological approach to optically control the electron dynamics of graphene nanoribbons, paving the way toward novel coherent electronic and quantum information processing devices in solid-state materials.
\end{abstract}

\maketitle

\section{Introduction}
Coherent electronics is an emerging field which aims to transfer quantum optical functionalities to the solid state domain, see e.g.~\cite{Higuchi2017,Heide2021,Kira2025}.
In contrast to this goal, standard electronics is based on voltage control of electron and hole currents~\cite{Hori1997, Cao2023}.
In this limit of electron dynamics, properties such as inter-band coherence, are negligible~\cite{Ashcroft1976, Kohn1957}.
However, coherent electronics utilizes quantum coherence as a central ingredient, such as superposition and entanglement, that can be used for a wide range of quantum technological applications~\cite{Dowling2003}.

Light-driven graphene nanoribbons are promising candidates for coherent electronics~\cite{Higuchi2017,Heide2021,Collado2025}, due to their large electro-optical response, high electron mobility, and the general robustness of graphene~\cite{CastroNeto2009,GarciaDeAbajo2014,Bonaccorso2010}.
In particular, armchair graphene nanoribbons exhibit an energy band gap in their electronic band structure that can be experimentally tuned by adjusting the width of the nanoribbon~\cite{CastroNeto2009,Zheng2007}.
These nanoribbons can be synthesized with atomic precision~\cite{Verzhbitskiy2016,Wang2021} and used to design electronic devices~\cite{Collado2025}.
Control of electron dynamics in graphene nanoribbons has been reported in Refs.~\cite{Jensen2013,Piskunow2014,Babajanov2014,Higuchi2017,Koppens2014,Wang2021,Singh2025}.

In this paper, we address a key challenge to generate \textit{coherent} electron dynamics in solids such as graphene nanoribbons.
In contrast to quantum systems with well-isolated eigenstates, graphene nanoribbons and solids in general, feature curved band structures.
Therefore, a coherent drive induces Rabi oscillations primarily for quasiparticles at momenta, at which the drive is on resonance with the band energy difference.
Quasiparticles close to resonance also respond to the drive, but oscillate at different frequencies, which introduces dephasing.
Consequently, coherent light does not produce collective coherent control of electron dynamics, which is a key requirement for the development of coherent electronics and quantum information technologies using solid-state systems.

\begin{figure}[!tb]
  \includegraphics[width=0.48\textwidth]{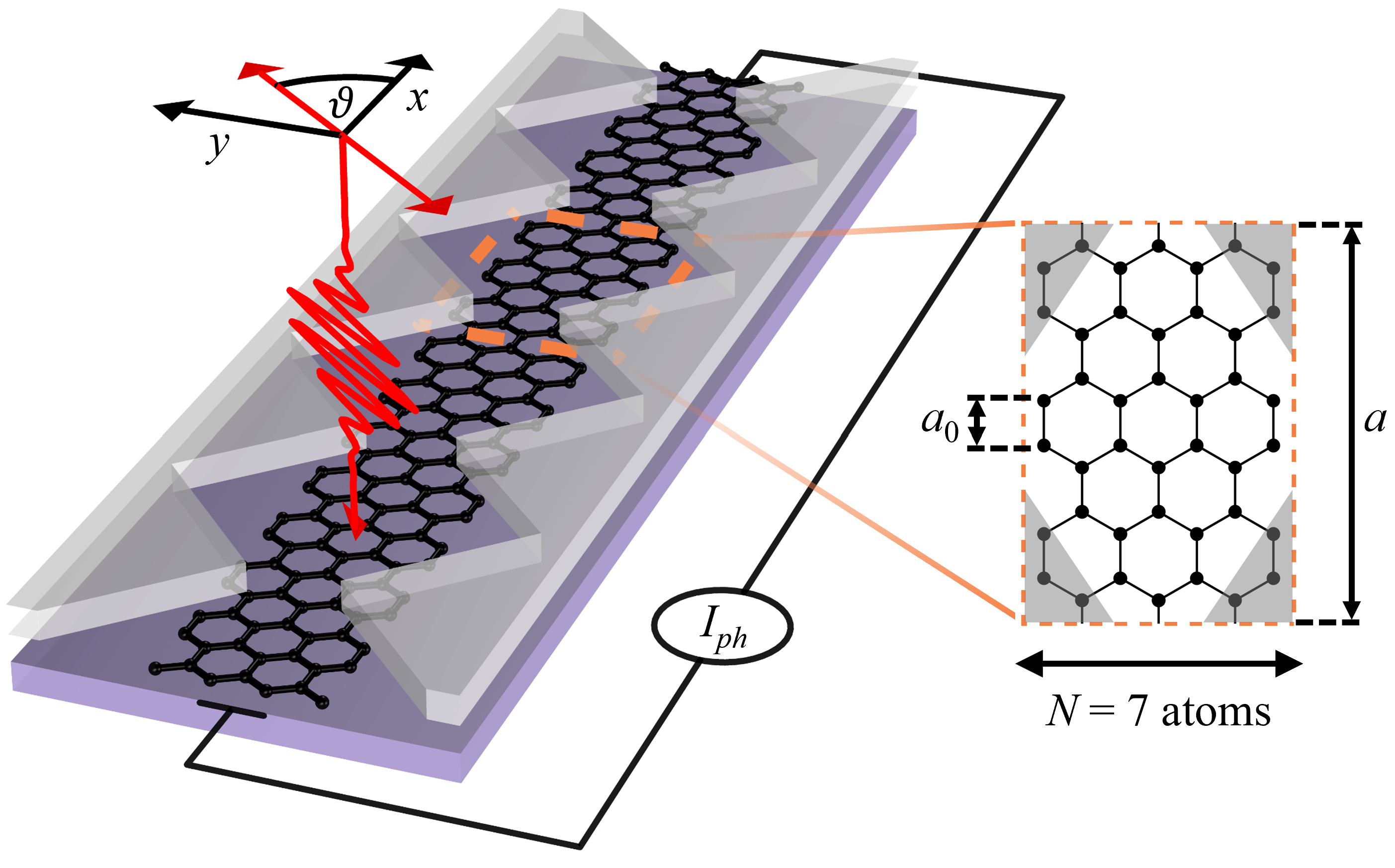}
  \caption{
  Patterned armchair graphene nanoribbon device. The periodically arranged gates (gray) introduce a one-dimensional superlattice, with lattice constant $a = \frac{3}{2}(N-1)a_0$, where $a_0 \approx 1.42\ \text{\AA}$ is the atom-atom distance and $N=7$ is the number of atoms along the finite direction of the nanoribbon. A zoom in of the unit cell of this superlattice is shown in the dashed orange box. The longitudinal photocurrent $I_{ph}$ can be coherently controlled by applying a laser pulse (red) which is linearly polarized at an optimal angle $\vartheta$.}
  \label{fig 1}
\end{figure}

To address this challenge, we propose an armchair graphene nanoribbon subjected to an external gate voltage pattern, as illustrated in Fig.~\ref{fig 1}, which enables flattening of the electronic band structure specifically for the valence and conduction band.
For this setup, we design an optimal electromagnetic driving pulse to produce collective and coherent Rabi oscillations of all quasiparticles in the valence band.
The combination of the optimized driving field and the gate voltages allows us to manipulate the photocurrent and electronic population between the valence band and conduction band while avoiding dephasing, thus enabling coherent electronics, and potentially quantum information processing.
We demonstrate the functionality of our methodology by simulating the unitary time evolution of the system for optimal pulses on a time scale of $10-100\ \si{fs}$, considering that dissipation can be neglected on such a time scale.

To illustrate our methodology, we show the electron dynamics for optimal $\pi/2$ and $\pi$ pulses and demonstrate that a second pulse can be used to reset the system to its initial state, recovering the equilibrium electronic distribution in a coherent manner.
Finally, given the tremendous progress in time and angle-resolved photoemission spectroscopy (trARPES)~\cite{Freericks2009, Sprinkle2009, Boschini2024,Merboldt2025}, we simulate trARPES of the system under the action of these pulses and conclude that this technique can be used to directly test our predictions experimentally.

This paper is organized as follows. In Sec.~\ref{sec II} we discuss how to engineer the band structure of the patterned nanoribbon to support quantum optical control.
In Sec.~\ref{sec III} we include the coupling to the electromagnetic field, which we optimize in Sec.~\ref{sec IV} to engineer a Rabi Hamiltonian.
We demonstrate in Sec.~\ref{sec V} coherent control of the electron dynamics of graphene nanoribbons.
In particular, we show that a $\pi$ pulse inverts the band population of the valence band and conduction band, and a $\pi/2$ pulse prepares a superposition state where an alternating photocurrent is generated.
Finally, in Sec.~\ref{sec VI}, we determine the electron distribution $n(k, \omega)$ which can be measured in time-resolved ARPES experiments.

\section{Band structure engineering using gate voltages}\label{sec II}
We propose a periodic array of voltage gates to control the electronic band structure of the system.
We show that for large gate voltages the low-energy bands become flat, which is the first step towards a coherent control of the system.

Specifically, we propose a periodic array of triangular-shaped control gates along both sides of an armchair graphene nanoribbon, as shown in Fig.~\ref{fig 1}.
These control gates introduce a superlattice for the electrons with a lattice constant $a = \frac{3}{2}(N-1)a_0$, where $a_0 \approx 1.42\ \text{\AA}$ is the atom-atom distance in graphene and $N$ is the number of atoms along the transverse direction of the nanoribbon, which defines its width.
For concreteness, we set $N=7$ for the remainder of the paper, although, we emphasize that our results also hold for larger nanoribbon widths.

The system is described by the tight-binding Hamiltonian
\begin{equation}\label{eqn 1}
  \hat{H} = -\sum_{\left<i,j\right>} |t| \hat{c}_{i}^\dagger \hat{c}_j +\sum_j U_j \hat{c}_{j}^\dagger \hat{c}_{j},
\end{equation}
where $\hat{c}_{j}^\dagger$ and $\hat{c}_{j}$ are the creation and annihilation operators for an electron on site $j$.
$t=-2.97\ \si{eV}$ is the tunneling energy between the nearest-neighbor atoms and $U_j$ is the on-site potential at site $j$ due to the gate voltages.
In the following we set $U_j$ to a constant value $U > 0$ for all sites near the control gates i. e. the gray regions in Fig.~\ref{fig 1}, and zero otherwise.
We write the Fourier-transformed Hamiltonian in a spinor representation as
\begin{equation}
  \hat{H} = \sum_k \hat{\Psi}_k^\dagger H_k \hat{\Psi}_k,
\end{equation}
where the spinor $\hat{\Psi}_k^\dagger = (\hat{c}_{k,1}^\dagger, \hat{c}_{k,2}^\dagger, \dots, \hat{c}_{k,M}^\dagger)$ contains the creation operators of electrons with momentum $k$ for each atom of the unit cell.
For $N=7$, there are $M=N(N-1)=42$ atoms in the unit cell, as can be seen in Fig.~\ref{fig 1}.
The elements of the matrix $H_k$ are given by
\begin{equation}\label{eqn 3} 
  H_{k,ij} = -|t| e^{ik(x_i - x_j)} \Theta(a_0 - |\boldsymbol{r}_{i} - \boldsymbol{r}_{j}|) + U_j \delta_{ij},
\end{equation}
where the Heaviside step function $\Theta(a_0 - |\boldsymbol{r}_{i} - \boldsymbol{r}_{j}|)$ ensures that only the nearest neighbor hopping terms are included, with $\Theta(0)=1$, and $\boldsymbol{r}_j = (x_j, y_j)$ denotes the position of the atom $j$.
We use the convention that the x- and y-directions correspond to the longitudinal and transversal directions, respectively, see Fig.~\ref{fig 1}.

\begin{figure}[t]
  \includegraphics[width=0.48\textwidth]{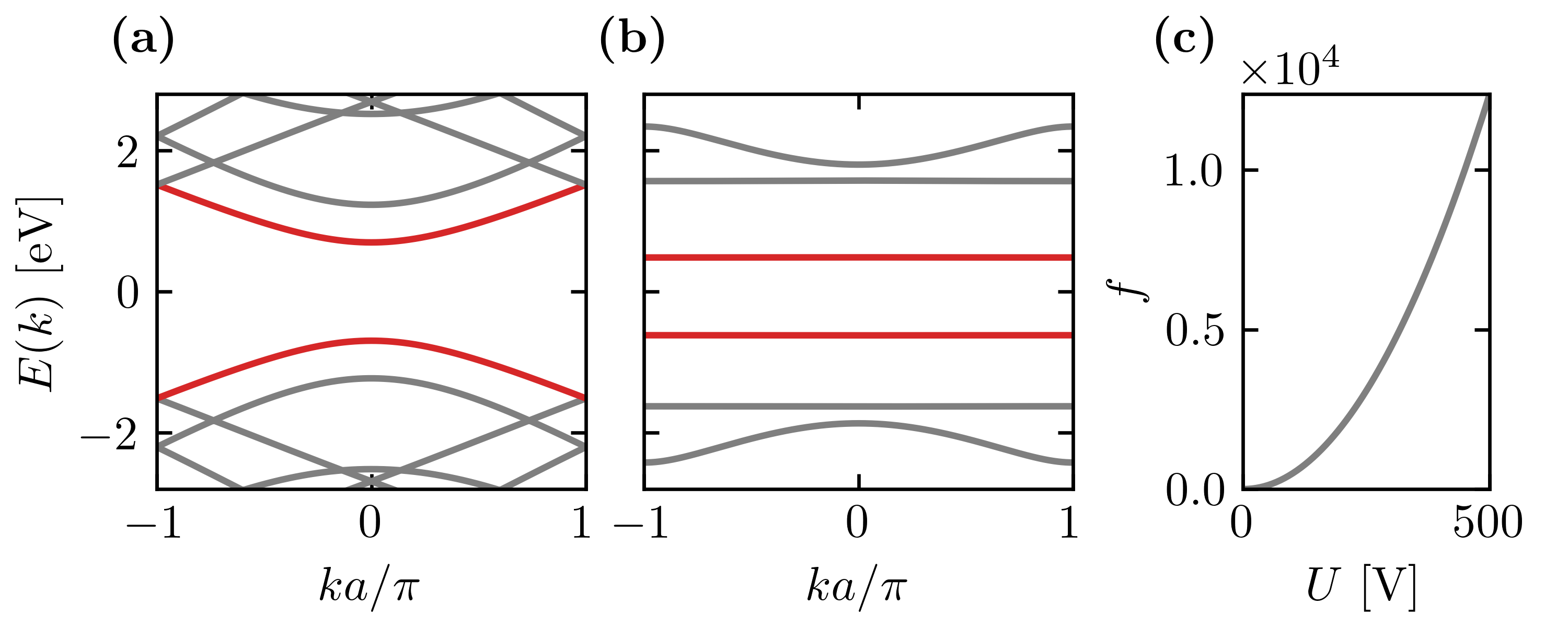}
  \hypertarget{fig:2c}{}
  \caption{
  Band structure for different gate voltages. Low-energy bands for $U = 0\ \si{V}$ (a) and $U = 100\ \si{V}$ (b). (c) The band flatness $f$ of the two lowest-energy bands (red), calculated by Eq.~(\ref{eqn 4}), as a function of the gate voltage.}
  \label{fig 2}
\end{figure}

We illustrate how the gate voltage pattern generates the desired flat band structure in Fig.~\ref{fig 2}, where we show the band structure associated with the Hamiltonian Eq.~(\ref{eqn 3}) for zero and high gate voltage, $U = 100\ \si{V}$.
As the gate voltage $U$ increases, the lowest-energy bands, plotted in red, flatten and the band gap decreases slightly.
For sufficiently high gate voltages, the band gap converges to $\Delta = 1.104\ \si{eV}$, see Appendix~\ref{appendix A}.
To quantify the effect of the gate voltage on the flattening of the band structure, we define the band flatness $f$ of the $m$th band with energy $\varepsilon_m(k)$ as the ratio between the minimal band gap $\Delta_m$ and the bandwidth $W_m$~\cite{Lee2016},
\begin{equation}\label{eqn 4}
  f = \frac{\Delta_m}{W_m} = \frac{min_k\{\varepsilon_m(k) - \varepsilon_{m-1}(k), \varepsilon_{m+1}(k) - \varepsilon_m(k)\}}{max_{k,k'}\{\varepsilon_m(k) - \varepsilon_m(k')\}}.
\end{equation}
The flatness of the lowest-energy bands displays approximately a quadratic dependence on the gate voltage, as shown in Fig.~\hyperlink{fig:2c}{2(c)}.
This indicates that it is possible to continuously tune the flatness of the bands by increasing the gate voltage with the proposed pattern.

\section{Coupling to the electromagnetic field}\label{sec III}
We now include the coupling to the electromagnetic field via the Peierls substitution~\cite{Peierls1933}, diagonalize the $42 \times 42$ Hamiltonian and reduce it to an effective $2 \times 2$ Hamiltonian by considering only the two lowest energy bands which can be flattened by increasing the gate voltage $U$.

In the presence of an electromagnetic field, the elements of the full Hamiltonian $H_k$ read
\begin{equation}
  H_{k,ij}(\boldsymbol{A}) = -|t| e^{i\phi_{ij}(\boldsymbol{A})} e^{ik(x_i - x_j)} \Theta(a_0 - |\boldsymbol{r}_{i} - \boldsymbol{r}_{j}|) + U_j \delta_{ij},
\end{equation}
where $\phi_{ij}(\boldsymbol{A}) = -\frac{e}{\hbar} \int_{\boldsymbol{r}_j}^{\boldsymbol{r}_{i}} \boldsymbol{A} d\boldsymbol{r}$~\cite{Peierls1933}, is the Peierls phase acquired by the electrons when tunneling from lattice site $j$ to lattice site $i$ in the presence of a vector potential $\boldsymbol{A}=(A_x, A_y)$.

We decompose $H_k$ into different contributions that account for the on-site potential and the six possible tunneling directions in the hexagonal structure of graphene.
This allows us to include the dependence on the vector potential as phase factors.
It reads
\begin{widetext}
\begin{eqnarray}\label{eqn 6}
  H_k(\boldsymbol{A}) &=& U h^{(0)} - |t| e^{2i\phi_x(\boldsymbol{A})} h^{(1)} - |t| e^{-2i\phi_x(\boldsymbol{A})} h^{(2)}
  - |t| e^{i(\phi_x(\boldsymbol{A}) + \phi_y(\boldsymbol{A}))} h^{(3)}\nonumber\\
  &&- |t| e^{-i(\phi_x(\boldsymbol{A}) + \phi_y(\boldsymbol{A}))} h^{(4)}
  - |t| e^{i(\phi_x(\boldsymbol{A}) - \phi_y(\boldsymbol{A}))} h^{(5)} - |t| e^{-i(\phi_x(\boldsymbol{A}) - \phi_y(\boldsymbol{A}))} h^{(6)},
\end{eqnarray}
\end{widetext}
where
\begin{eqnarray} \label{eqn 7}
  \phi_x(\boldsymbol{A}) &=& -\frac{ea_0}{2\hbar} A_x + \frac{ka_0}{2}, \\
  \phi_y(\boldsymbol{A}) &=& -\frac{\sqrt{3}ea_0}{2\hbar} A_y
\end{eqnarray}
are the phases acquired for electrons by tunneling in x- and y-direction, respectively.
The matrices $h^{(\lambda)}$ with $\lambda=0,\dots,6$ in Eq.~(\ref{eqn 6}) are independent of the vector potential and derive from which sites are connected via tunneling.
The matrix element $h_{ij}^{(\lambda)}$ is one if the sites $i$ and $j$ are nearest neighbors and oriented along the direction corresponding to $\lambda$, and zero otherwise.

We now apply the $k$-dependent transformation
\begin{eqnarray}
    \widetilde{\Psi}_k &=& U_k \Psi_k, \\
    \widetilde{H}_k &=& U_k H_k U_k^\dagger \label{eqn 10},
\end{eqnarray}
where the matrix $U_k$ contains the eigenstates of $H_k$ for zero vector potential.
Thus, after applying this transformation, the Hamiltonian $\widetilde{H}_k$ becomes diagonal for zero vector potential, and nondiagonal for any finite vector potential.

We reduce $\widetilde{H}_k$ to a $2\times2$ effective Hamiltonian $H_{k}^{\mathrm{eff}}(\boldsymbol{A})$ by projecting onto the two lowest-energy bands, which are indicated in red in Fig.~\ref{fig 2}.
The dependence of this effective Hamiltonian on the vector potential can be derived from Eq.~(\ref{eqn 6}), where the matrices $h^{(\lambda)}$ are replaced by effective $2\times2$ matrices $\widetilde{h}^{(\lambda)^{\mathrm{eff}}}$, which are first transformed as $\widetilde{h}^{(\lambda)} = U_k h^{(\lambda)} U_k^\dagger$, and then reduced to the effective $2\times2$ subspace.
The full expression for $H_{k}^{\mathrm{eff}}(\boldsymbol{A})$ and useful symmetries are shown in the Appendix~\ref{appendix B}.
In the next sections it will be useful to express $H_{k}^{\mathrm{eff}}(\boldsymbol{A})$ in terms of the Pauli matrices.
Therefore, we decompose it as
\begin{eqnarray}\label{eqn 11}
  H_{k}^{\mathrm{eff}}(\boldsymbol{A}) &=& H_{k,1\kern-0.22em\text{l}}(\boldsymbol{A}) 1\kern-0.25em\text{l} + H_{k,x}(\boldsymbol{A}) \sigma_x \nonumber\\
  &+& H_{k,y}(\boldsymbol{A}) \sigma_y + H_{k,z}(\boldsymbol{A}) \sigma_z.
\end{eqnarray}
Note that the transformation Eq.~(\ref{eqn 10}) imprints global phases on the off-diagonal elements of $\widetilde{H}_k$ depending on the phases of the eigenstates of $H_k(\boldsymbol{A}=0)$.
Therefore, $H_{k,x}$ and $H_{k,y}$ depend on the choice of these phases.
Throughout the paper, we choose a global phase such that $H_{k,x}$ always remains zero and only $H_{k,y}$ is finite for a finite vector potential.

\section{Engineering a Rabi Hamiltonian}\label{sec IV}
Having derived the effective two-band Hamiltonian $H_{k}^{\mathrm{eff}}(\boldsymbol{A})$, we now describe how to implement approximately the Rabi Hamiltonian given by
\begin{equation} \label{eqn 12}
  H_{\mathrm{Rabi}} = \begin{pmatrix}
    \Delta/2 & -i\hbar\Omega(t) \sin(\omega_{dr} t) \\
    i\hbar\Omega(t) \sin(\omega_{dr} t) & -\Delta/2
  \end{pmatrix},
\end{equation}
where $\Delta$ is the band gap, $\omega_{dr}$ is the frequency of the driving field, and $\Omega(t)$ is the Rabi frequency.
We note that Eq.~(\ref{eqn 12}) recovers the standard representation of the Rabi oscillations around the y-axis of the Bloch sphere by taking the rotating-wave approximation.
We implement Eq.~(\ref{eqn 12}) by using linearly polarized light.
We write the linear relation between the vector potential $\boldsymbol{A} = (A_x, A_y)$ of the light pulse, and the Rabi term in Eq.~(\ref{eqn 12}) as
\begin{eqnarray}
  A_x = 2 A_0 \cos(\vartheta) \hbar \Omega(t) \sin(\omega_{dr}t), \label{eqn 13}\\
  A_y = 2 A_0 \sin(\vartheta) \hbar \Omega(t) \sin(\omega_{dr}t), \label{eqn 14}
\end{eqnarray}
where $\omega_{dr}$ is the driving frequency, and $\vartheta$ is the polarization angle of the linearly polarized light, and $A_0$ is a proportionality constant.
We note that a natural representation of the Rabi frequency is $\Omega \sim \boldsymbol{d} \cdot \boldsymbol{E}$, where $\boldsymbol{d}$ is the dipole moment of the interband transition and $\boldsymbol{E}$ is the electric field, which is related to the vector potential via $\boldsymbol{A} \sim \boldsymbol{E}/\omega_{dr}$.
Therefore, the inverse of $A_0$ provides the effective dipole moment of the interband transition.
In the following, we consider on-resonance driving, i.e., we fix the driving frequency to match the band gap of the patterned graphene nanoribbon, i.e., $\hbar\omega_{dr}=\Delta$.
Below, we determine the optimal choices for $A_0$ and $\theta$, to implement the Rabi Hamiltonian., for $k=0$. Then we confirm that the same set of parameters is optimal for all $k$.

As the first step, we calculate the optimal polarization angle by imposing that the diagonal elements of the effective Hamiltonian $H_{k}^{\mathrm{eff}}$ [Eq.~(\ref{eqn 11})] are approximately time independent.
Since the time dependence of the Hamiltonian derives from the vector potential, we demand that the following two conditions
\begin{eqnarray} \label{eqn 15}
  H_{k,1\kern-0.22em\text{l}}(A_x, A_y) = H_{k,1\kern-0.22em\text{l}}(0, 0)
\end{eqnarray}
and
\begin{eqnarray} \label{eqn 16}
  H_{k,z}(A_x, A_y) = H_{k,z}(0, 0)
\end{eqnarray} are fulfilled.
In Fig.~\hyperlink{fig:3a}{3(a)} we plot $\delta H_{k,1\kern-0.22em\text{l}}=|H_{k,1\kern-0.22em\text{l}}(A_x, A_y) - H_{k,1\kern-0.22em\text{l}}(0,0)|$ for $k=0$ and $U = 100\ \si{V}$, as a function of $A_x$ and $A_y$.
We observe that $\delta H_{0,1\kern-0.22em\text{l}}\approx0$ for sufficiently small values of $A_x$ and $A_y$, independently of the polarization angle and the gate voltage $U$ assuming relatively high gate voltages.
Thus, for $|\boldsymbol{A}| \ll \frac{2\hbar}{ea_0}$, Eq.~(\ref{eqn 15}) is approximately satisfied.
This result holds for all sufficiently large values of $U$.

\begin{figure}
  \includegraphics[width=0.48\textwidth]{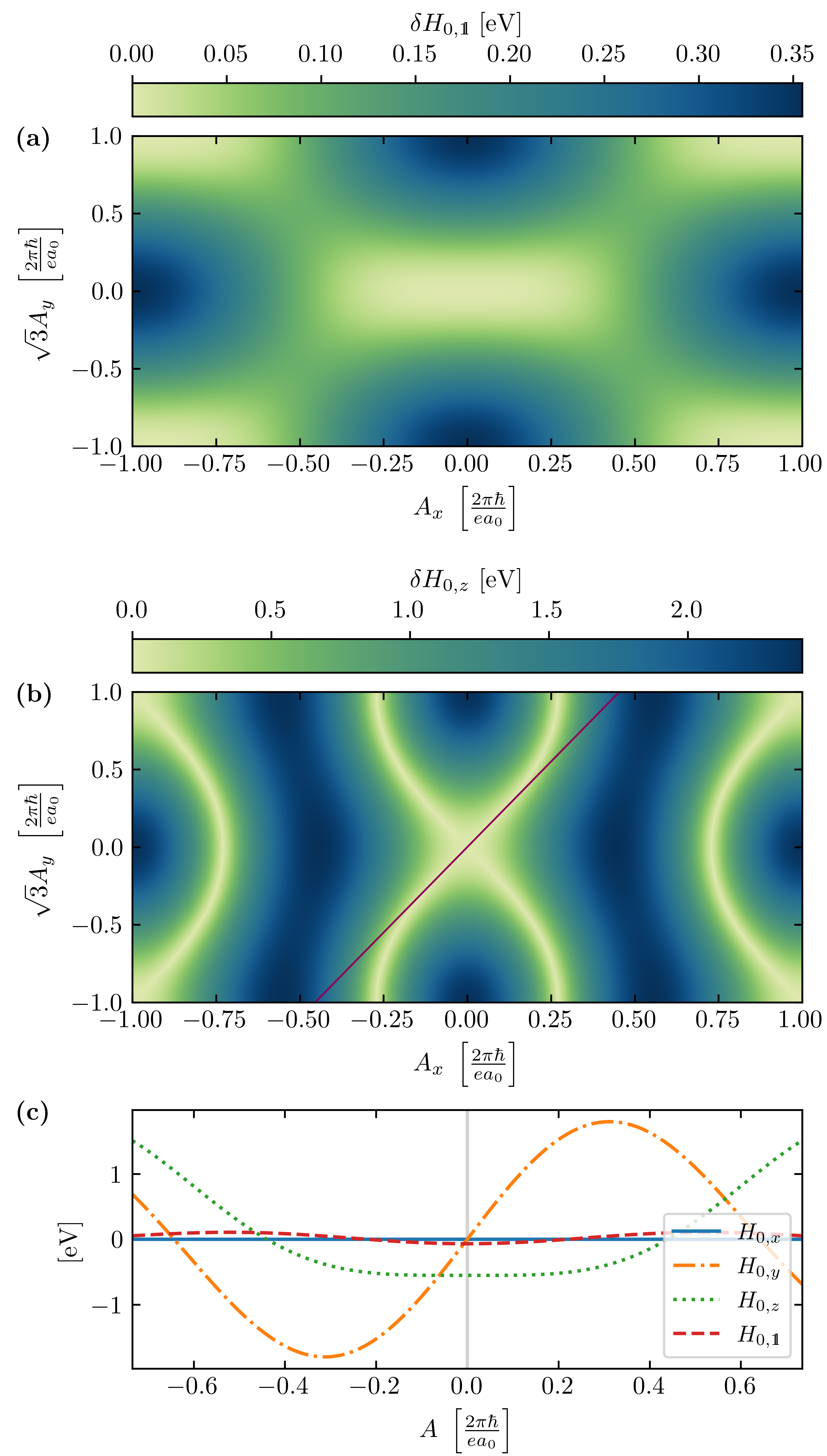}
  \hypertarget{fig:3a}{}
  \hypertarget{fig:3b}{}
  \hypertarget{fig:3c}{}
  \caption{Dependence of the components of the effective Hamiltonian on the vector potential for $k=0$. In panel (a) we show $\delta H_{0,1\kern-0.22em\text{l}}$ and in panel (b) we show $\delta H_{0,z}$. The purple diagonal line indicates the optimal polarization angle Eq.~(\ref{eqn 17}). (c) Components of the effective Hamiltonian for the optimal polarization angle and $k=0$. For small vector potentials $H_{0,y}$ is linear, $H_{0,z}$ and $H_{0,1\kern-0.22em\text{l}}$ are constant and $H_{0,x}$ is zero. The gate voltage is $U=100\ \si{V}$.}
  \label{fig 3} 
\end{figure}

\begin{figure}
  \includegraphics[width=0.48\textwidth]{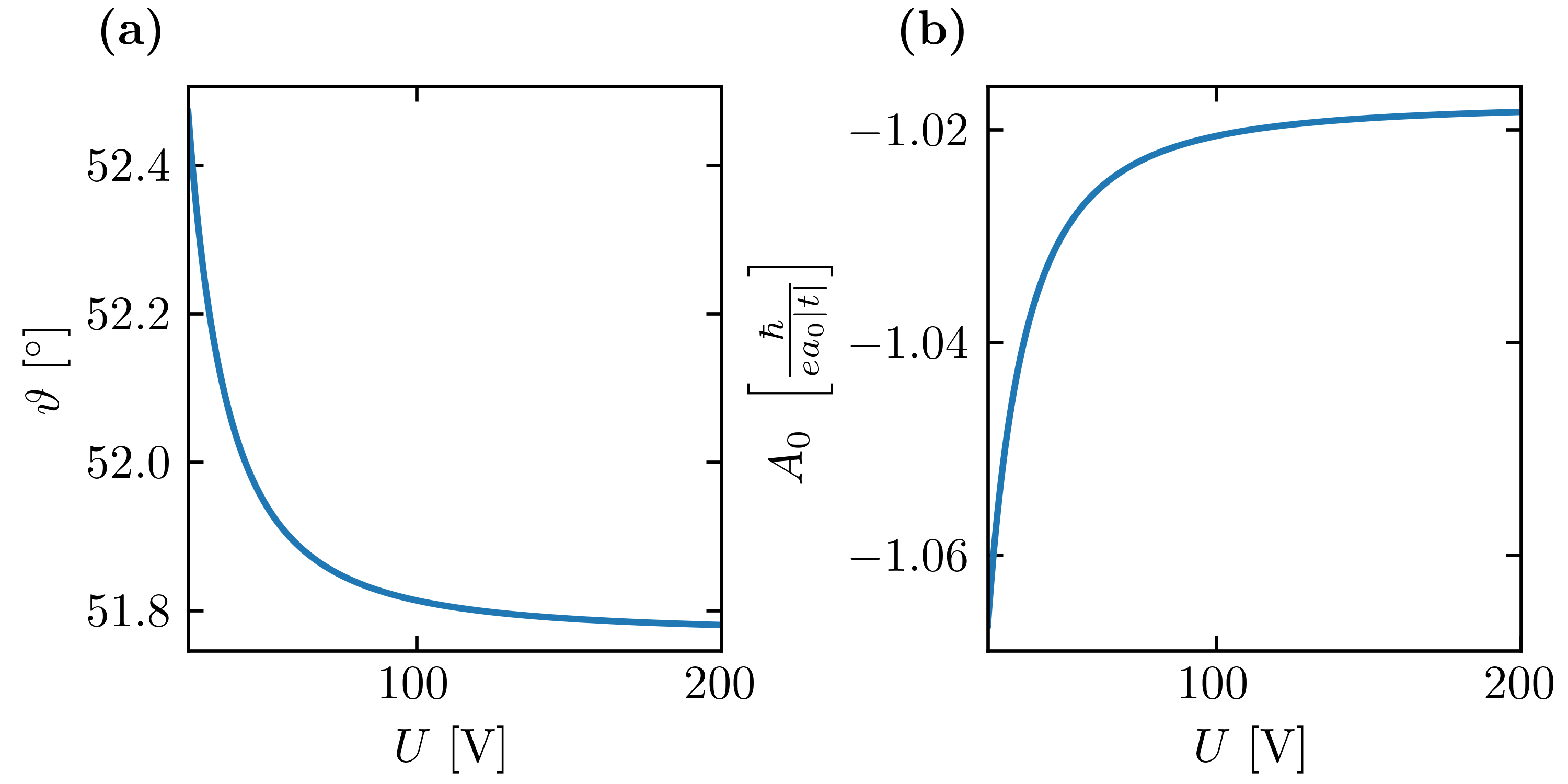}
  \hypertarget{fig:4a}{}
  \hypertarget{fig:4b}{}
  \caption{Dependence of the optimal polarization angle (a) and the optimal factor $A_0$ (b) on the gate voltage.}
  \label{fig 4}
\end{figure}

In Fig.~\hyperlink{fig:3b}{3(b)} we plot $\delta H_{k,z}=|H_{k,z}(A_x, A_y) - H_{k,z}(0,0)|$ for $k=0$ and a gate voltage $U = 100\ \si{V}$.
$\delta H_{0,z}$ vanishes along the direction given by an optimal polarization angle $\vartheta = 51.77^\circ$, which we indicate with a purple diagonal line. 
We indeed demonstrate that, for $k=0$, Eq.~(\ref{eqn 16}) can be satisfied for a linear polarization angle
\begin{equation}\label{eqn 17}
  \vartheta = \arctan\left(\frac{\gamma}{\sqrt{3}}\right)
\end{equation}
with
\begin{equation}
  \gamma = \sqrt{-\left(1 + 2\ \frac{\widetilde{h}_{00}^{(1)^{\mathrm{eff}}} - \widetilde{h}_{11}^{(1)^{\mathrm{eff}}}}{\widetilde{h}_{00}^{(3)^{\mathrm{eff}}} - \widetilde{h}_{11}^{(3)^{\mathrm{eff}}}}\right)}.
\end{equation}
The detailed derivation of this analytical result can be found in the Appendix~\ref{appendix C}.
This indicates a dependence of the optimal polarization angle $\vartheta$ on the applied gate voltage $U$ via the matrices $\widetilde{h}^{(\lambda)^{\mathrm{eff}}}$, as we show below.

Thus, the condition Eq.~(\ref{eqn 16}) is fulfilled for small vector potentials $|\boldsymbol{A}| \ll \frac{2\hbar}{ea_0}$ with the optimal polarization angle $\vartheta$ given by Eq.~(\ref{eqn 17}).

Next, we determine the amplitude factor $A_0$ to engineer the Rabi Hamiltonian in Eq.~(\ref{eqn 12}).
Crucially, for the optimal polarization angle and small vector potentials, there is a linear dependence of the off-diagonal elements of the effective Hamiltonian on the vector potential.
This can be seen in Fig.~\hyperlink{fig:3c}{3(c)}, where we plot the components of the effective Hamiltonian in the Pauli basis for $k=0$.
As depicted, $H_{0,x}$ is zero, while $H_{0,y}$ shows a linear dependence for small vector potentials.
The slope of the linear dependence of $H_{0,y}$ determines the response of the Hamiltonian to the driving field, i.e., the Rabi oscillations at $k=0$.
We account for this slope by choosing the factor $A_0$ such that the off-diagonal elements of the effective Hamiltonian and the specific Rabi Hamiltonian that we want to implement coincide.
It is
\begin{equation} \label{eqn 19}
  H^{\mathrm{eff}}_{k,01} = -i\hbar \Omega(t) \sin(\omega_{dr}t)
\end{equation}
for $k=0$.
We insert the driving field Eq.~(\ref{eqn 13}) and Eq.~(\ref{eqn 14}) into $H^{\mathrm{eff}}_{k,01}$ to find that, in the limit of weak driving $|\boldsymbol{A}| \ll \frac{2\hbar}{ea_0}$, the factor $A_0$ has to be
\begin{equation}\label{eqn 20}
  A_0 = -\frac{\hbar}{4ea_0|t|} \frac{1}{\cos(\vartheta)}\frac{1}{\widetilde{h}_{01}^{(1)^{\mathrm{eff}}}+\widetilde{h}_{01}^{(3)^{\mathrm{eff}}}}
\end{equation}
to fulfill Eq.~(\ref{eqn 19}).
For a detailed derivation of this analytical result, we refer the reader to Appendix~\ref{appendix D}.
For completeness, we also plot $H_{k,1\kern-0.22em\text{l}}$ and $H_{k,z}$ as a function of the vector potential for the optimal polarization angle and $k=0$ in Fig.~\hyperlink{fig:3c}{3(c)}.
They both show a flat dependence for small vector potentials, as expected from the conditions Eq.~(\ref{eqn 15}) and Eq.~(\ref{eqn 16}). 

By numerically evaluating Eq.~(\ref{eqn 17}) and Eq.~(\ref{eqn 20}) as a function of the gate voltage $U$, we find that the optimal polarization angle and optimal factor $A_0$ converge to $\vartheta = 51.77^\circ$ and $A_0 = -1.018\ \frac{\hbar}{ea_0|t|}$, respectively, for a sufficiently high gate voltage $U$, as shown in Fig.~\ref{fig 4}.

\section{Coherent control of band populations and photocurrent generations}\label{sec V}
To overcome dephasing and induce collective and coherent Rabi oscillations, it is crucial that the driving field is optimized not only for zero momentum $k=0$, but also for all other momenta $k$.
We study the electron dynamics for different $k$ states generated by the effective Hamiltonian Eq.~(\ref{eqn 11}) with the optimal driving field that we obtained for $k=0$ and different gate voltages.
As we show below, by increasing the gate voltages $U$, the optimal driving field for $k=0$ is also optimal for all momenta.

To illustrate coherent control of electronics in our system, we consider a $\pi$ pulse driving field.
We use a spinor representation for every state $k$.
In this representation, the states rotate on a Bloch sphere under the action of the driving field.
In particular, for an ideal $\pi$ pulse, the z-component of the different Bloch vectors $\left<\sigma_z\right>_k$ changes from $-1$ to $1$ due to the pulse.
In Fig.~\hyperlink{fig:5a}{5(a)}, we plot $\left<\sigma_z\right>_k$ after the optimal $\pi$ pulse for $k=0$ for different gate voltages $U$.
$\left<\sigma_z\right>_k=1$ ($\left<\sigma_z\right>_k=-1$) corresponds to a total occupation of the conduction (valence) band at momentum $k$.
For $k=0$, the light pulse implements a $\pi$ pulse by construction.
In Fig.~\hyperlink{fig:5a}{5(a)}, we show that the higher the gate voltage $U$, i.e., the flatter the bands are, the better the light pulse implements a $\pi$ pulse for all momenta, resulting in complete band inversion.
For gate voltages larger than $U=100\ \si{V}$, the upper band is populated at each momentum with an fidelity of $99.8\ \%$, so the bands are almost perfectly inverted.

Following the same strategy of band flattening, using the optimal $\pi/2$ pulse for $k=0$ and high gate voltages, we show that all Bloch vectors rotate from the south pole to the equator of the Bloch sphere, resulting in $\left<\sigma_z\right>_k=0$ right after the pulse for almost all momenta $k$ as shown in Fig.~\hyperlink{fig:5b}{5(b)}.
This optimal pulse generates a superposition state and alternating photocurrents as we demonstrate below.

\begin{figure}[!tb]
  \includegraphics[width=0.48\textwidth]{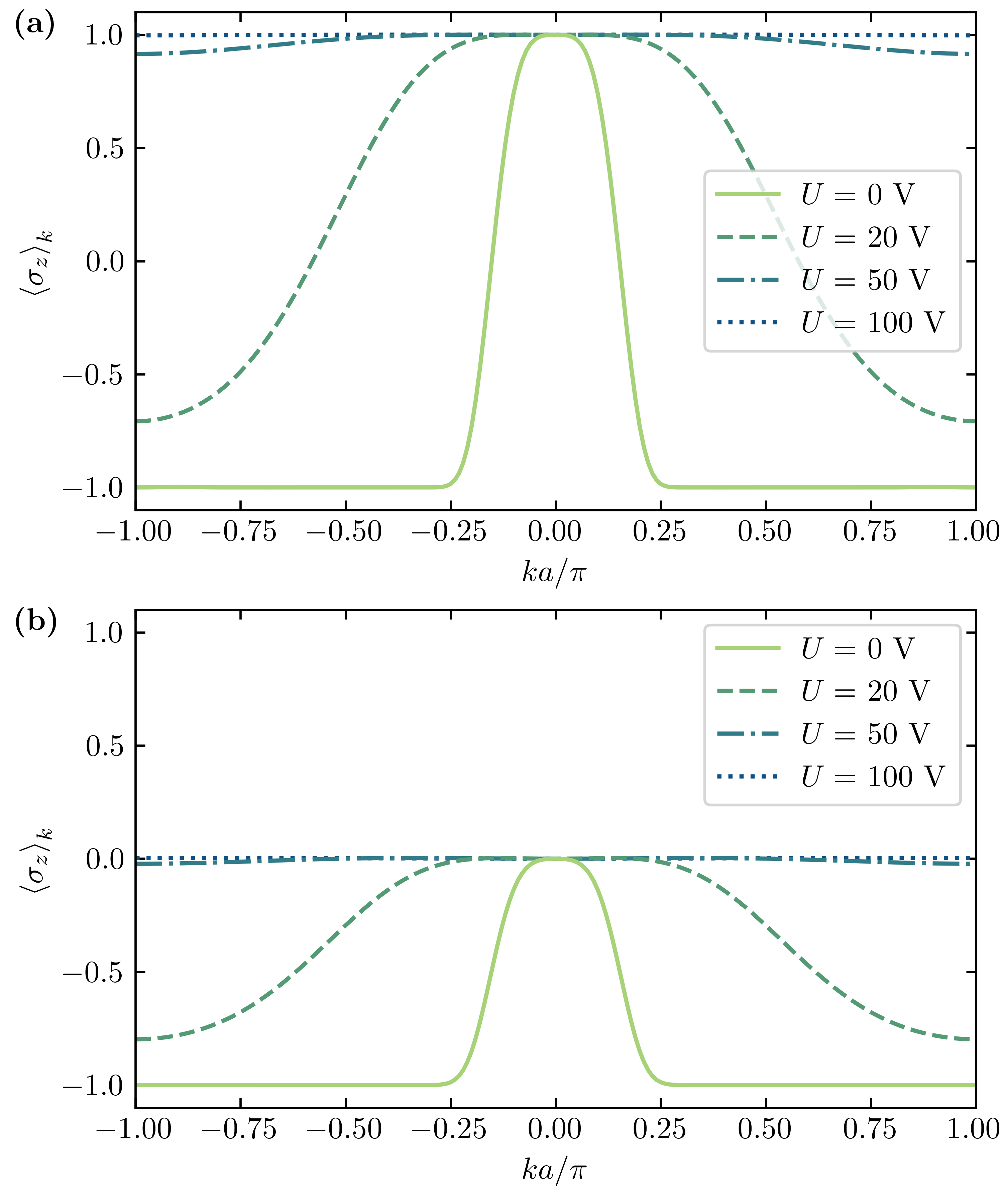}
  \caption{z-component of the Bloch vectors, $\left<\sigma_z\right>_k$, as a function of $k$ after applying a $\pi$ pulse (a) and after applying a $\pi/2$ pulse (b). For the simulation we use a pulse length of $\tau = 20\ \si{fs}$ (FWHM) and the optimal polarization angles and factors $A_0$ determined by Eq.~(\ref{eqn 17}) and Eq.~(\ref{eqn 20}), respectively. The driving frequencies are tuned to the band gaps which are around $\Delta \approx 1.11\ \si{eV}$ depending on the gate voltage, see Appendix~\ref{appendix A}.}
  \label{fig 5}
\end{figure}

\begin{figure}[!tb]
  \hypertarget{fig:6a}{}
  \hypertarget{fig:6b}{}
  \includegraphics[width=0.48\textwidth]{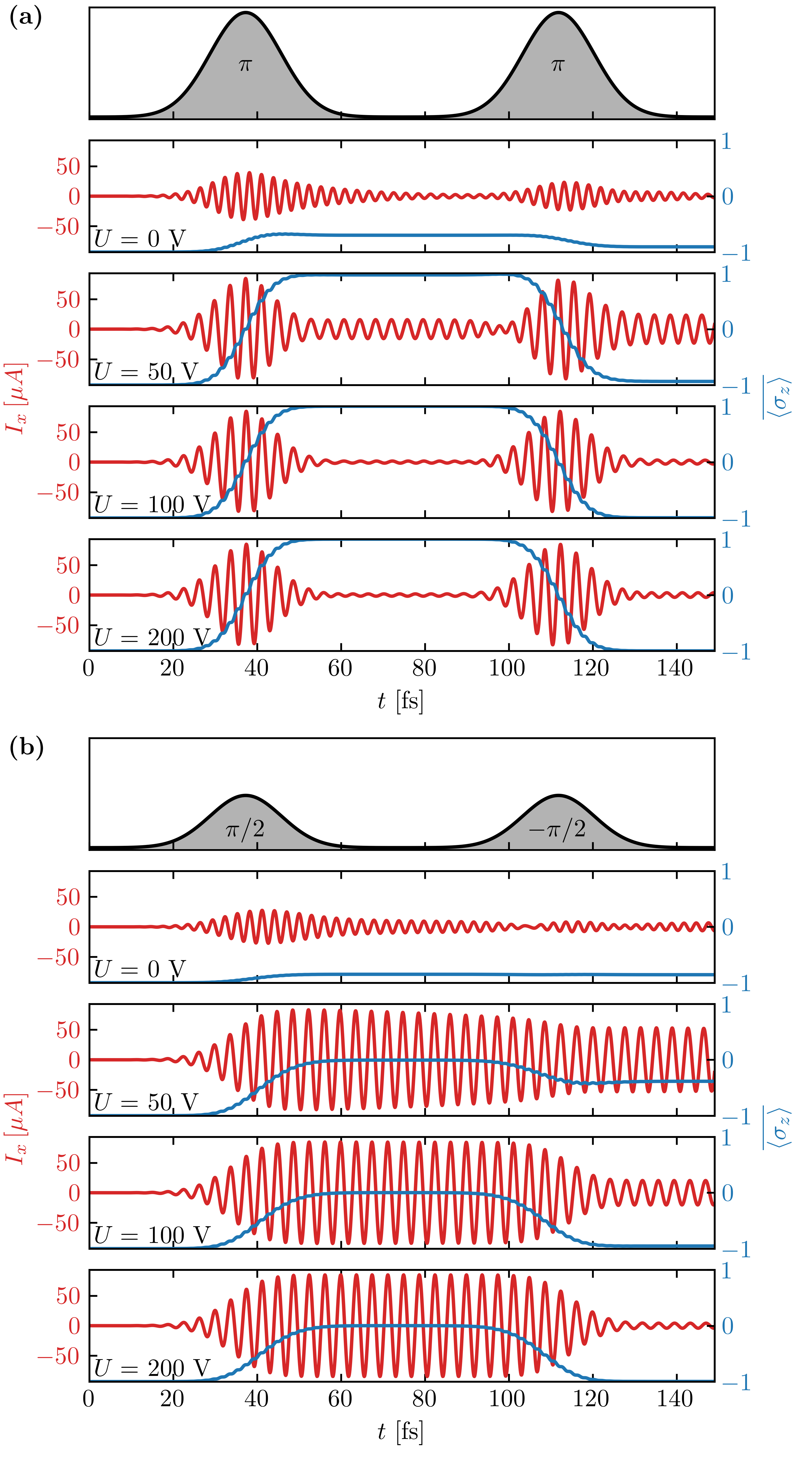}
  \caption{Photocurrent and band population dynamics for a sequence of two $\pi$ pulses (a) and a sequence of two $\pi/2$ pulses (b). In both cases, the upper panel shows the Gaussian envelope function of the pulses in arbitrary units. Below, we show the time dependence of the photocurrent (red) and of the z-component of the Bloch vectors averaged over all momentum states (blue) for different gate voltages $U$. We use a fixed pulse width $\tau = 20\ \si{fs}$ (FWHM). The optimal polarization and optimal factor $A_0$ for each gate voltage are determined by Eq.~(\ref{eqn 17}) and Eq.~(\ref{eqn 20}), respectively. The driving frequency is tuned to the band gap which is around $\Delta \approx 1.11\ \si{eV}$ depending on the gate voltage, see Appendix~\ref{appendix A}.}
  \label{fig 6}
\end{figure}

The $\pi/2$ light pulse that we have designed and applied in Fig.~\ref{fig 5} generates a photocurrent along the longitudinal direction of the nanoribbon, which we calculate via
\begin{equation}
  I_x = \sum_k \left\langle \frac{\partial H_{k}^{\mathrm{eff}}}{\partial A_x}\right\rangle.
\end{equation}
To quantify the degree of coherence of our pulse protocols, we plot the time evolution of the photocurrent for a sequence of two $\pi$ pulses in Fig.~\hyperlink{fig:6a}{6(a)} and for a sequence of two $\pi/2$ pulses in Fig.~\hyperlink{fig:6b}{6(b)} for different gate voltages.
The second pulse in each protocol is designed such that the initial state is recovered if no dephasing is induced during the action of the driving fields.
We also plot the time evolution of the z-component of the Bloch vectors averaged over all momenta $k$, $\overline{\left<\sigma_z\right>}$.
The simulations start at equilibrium $(t=0)$ with a fully occupied valence band.
Thus, $\left<\sigma_z\right>_k=-1$ for all momenta $k$ and $\overline{\left<\sigma_z\right>}=-1$.

In an ideal $\pi$ pulse scheme, the first $\pi$ pulse transfers all Bloch vectors from the bottom of the Bloch sphere to the top of the Bloch sphere producing $\left<\sigma_z\right>_k=1$.
This does not result in a photocurrent but instead results in an inversion of the band populations.
The second $\pi$ pulse transfers all Bloch vectors back to the bottom of the Bloch sphere, which is the initial state.
This ideal picture is valid for gate voltages above $U \sim 100\ \si{V}$ whereas for smaller gate voltages a residual oscillating photocurrent remains after the second $\pi$ pulse as a consequence of dephasing.
This can be seen, for example, for the $U=50\ \si{V}$ case at $t>120\ \si{fs}$.

In the $\pi/2$ pulse scheme, all Bloch vectors are driven from the south pole of the Bloch sphere to the equator, resulting in $\left<\sigma_z\right>_k=0$ for almost all momenta $k$ assuming a high gate voltage $U$, see Fig.~\hyperlink{fig:6b}{6(b)}.
After the $\pi/2$ pulse, the Bloch vectors rotate along the equator, around the z axis of the Bloch sphere, resulting in a strong alternating current because all Bloch vectors advance in phase and their contributions to the current add up constructively.
The final state is a superposition of each valence and conduction momentum state.
The first $\pi/2$ pulse drives this state preparation almost perfectly for $U\geq100\ \si{V}$ for all Bloch vectors, such that $\left<\sigma_z\right>_k\simeq0$ for almost all momenta $k$ and thus $\overline{\left<\sigma_z\right>}\simeq0$.
The second $\pi/2$ pulse has the opposite sign and therefore, all Bloch vectors are approximately transferred back to the bottom of the Bloch sphere resulting in a vanishing photocurrent.
However, we notice that in general the second $\pi/2$ pulse does not perfectly revert the dynamics, and higher gate voltages are needed to obtain a zero current after the two-pulse sequence. 
The reason for the different performance of the first and second $\pi/2$ pulse is the remaining dispersion, which leads to dephasing of the Bloch vectors. 
For $U=200\ \si{V}$ the rotation of $\pi/2$ back to the initial state is achieved and the photocurrent vanishes similarly to the ideal case scenario.
This effect does not appear in the case of $\pi$ pulses in Fig.~\hyperlink{fig:6a}{6(a)}, because a phase shift between two $\pi$ pulses does not affect the action of the second pulse.

These results show the remarkable coherent control of electron dynamics in this solid-state platform because of the flatness of the bands and the use of optimized driving fields.

\begin{figure}[!tb]
  \includegraphics[width=0.48\textwidth]{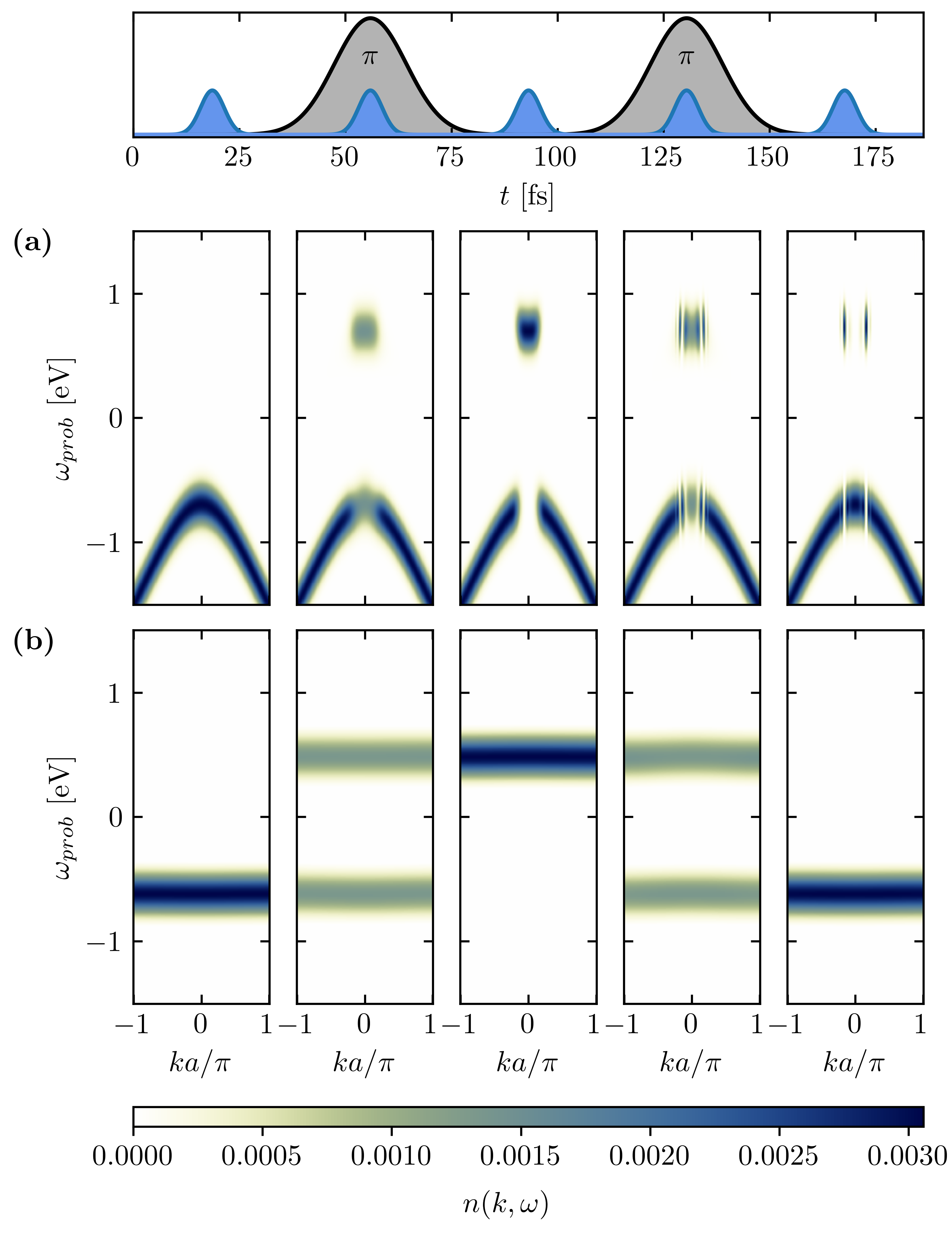}
  \caption{The electron distribution at five different times during a sequence of two $\pi$ pulses. The upper panel displays the two $\pi$ pulses (gray) and the five probing pulses (blue) with rescaled amplitudes as an illustration. Below, in panels (a) and (b), we show the electron distributions $n(k,\omega)$ for the five probing pulses at different times which are calculated using Eq.~(\ref{eqn 22}). We show the results for a gate voltage of $U=0\ \si{V}$ (a) and $U=100\ \si{V}$ (b), a driving pulse length of $20\ \si{fs}$ (FWHM) and a probing pulse length of $10\ \si{fs}$ (FWHM). The optimal polarization for each gate voltage is determined by Eq.~(\ref{eqn 17}), which is $\vartheta = 49.89^\circ$ for $U = 0\ \si{V}$ and $\vartheta =51.81^\circ$ for $U = 100\ \si{V}$. The optimal factor $A_0$ is determined by Eq.~(\ref{eqn 20}), which is $A_0 = -0.561\ \frac{\hbar}{ea_0|t|}$ for $U = 0\ \si{V}$ and $A_0 = -1.021\ \frac{\hbar}{ea_0|t|}$ for $U = 100\ \si{V}$. The driving frequency is tuned to the band gap, which is $\Delta = 1.394\ \si{eV}$ for $U = 0\ \si{V}$ and $\Delta =1.106\ \si{eV}$ for $U = 100\ \si{V}$.}
  \label{fig 7}
\end{figure}

\section{Time-resolved ARPES}\label{sec VI}
In this section, we determine the signature of the electron distribution $n(k, \omega)$ in a time-resolved ARPES (trARPES) experiment, for the two $\pi$ pulse protocol.
This corresponds to measurement of the electron distribution that is Fourier broadened due to the finite pulse length.
We calculate the trARPES signature of the electron distribution as~\cite{Freericks2009, Broers2022, Nuske2020, Broers2021}
\begin{equation} \label{eqn 22}
  n(k, \omega) = \underset{t_0}{\overset{t_{f}}{\int}} dt_1 \underset{t_0}{\overset{t_{f}}{\int}} dt_2\ \frac{s(t_1) s(t_2) e^{-i\omega(t_2 - t_1)}}{4(t_{f}-t_0)^2}\mathcal{G}(k,t_2, t_1),
\end{equation}
where $\omega$ is the probing frequency and 
\begin{equation}
  \mathcal{G}(k,t_2, t_1) = \left\langle \hat{c}_k^\dagger(t_2) \hat{c}_k(t_1) \right\rangle
\end{equation} is the single-particle correlation function. 
$s(t)$ denotes the probing pulse envelope which we take as
\begin{equation}
  s(t) = \exp\left(-4\ln{2}\frac{(t-t_{prob})^2}{\tau_{prob}^2}\right),
\end{equation}
where $t_{prob}$ is the center of the probing pulse and $\tau_{prob}$ is the pulse length (FWHM).
We consider the integration limits as $t_0=0$ and $t_f=186.4\ \si{fs}$ which is the entire time window of our simulation.
As mentioned, the time-resolved measurement broadens the electron distribution on the scale of $1/\tau_{probe}$.
In the following, we refer to $n(k, \omega)$ as the electron distribution, for simplicity.

In Fig.~\ref{fig 7} we show the electron distribution at five different times during a sequence of two $\pi$ pulses for zero and high gate voltage.
For the case of zero gate voltage, the electron distribution dynamics show how the first pulse excites only the electrons near the resonance at $k=0$ and the second pulse does not completely revert these excitations to the initial state due to the curved band structure and the resulting dephasing between different momenta $k$ in Fig.~\hyperlink{fig:7a}{7(a)}.
However, in the case of a high gate voltage, the first $\pi$ pulse perfectly inverts the band population of the flat bands as can be seen by probing the electron distribution right after the pulse, see center panel in Fig.~\hyperlink{fig:7b}{7(b)}.
The second $\pi$ pulse inverts the population back to the initial state where the valence band is completely filled as observed in the trARPES signal after the second pulse, see right panel in Fig.~\hyperlink{fig:7b}{7(b)}.
The second and fourth probing pulses are applied at the same times as the $\pi$ pulses, so half-way through the $\pi$ pulses.
Therefore, the electron distributions corresponding to these probing pulses are equivalent to those obtained after applying a $\pi/2$ pulse where the band population is equally distributed between the valence and conduction bands.

This shows how trAPPES constitutes a suitable and direct experimental technique to test our predictions on coherent control of electron dynamics in the setup we propose.

\section{Conclusion}
In conclusion, we have demonstrated that the band populations of armchair graphene nanoribbons can be coherently controlled by a combination of optimal linearly polarized driving pulses and electrostatic gates along the nanoribbon.
The key principle is that gate voltages are used to flatten the band structure of the nanoribbon such that the driving pulses act equally for all momenta $k$.
Because of the nontrivial structure of the eigenstates for the patterned graphene nanoribbon, the driving pulse has to be optimized depending on the applied gate voltage $U$.
We have designed the optimal driving field parameters to engineer any collective Rabi oscillations $\Omega(t)$, i.e., the optimal polarization and amplitude factor to rotate the electronic states to an arbitrary angle on the Bloch sphere.
We have shown that the higher the gate voltages the flatter are the valence and conduction band and the higher is the coherence between the bands, in response to the optimal driving field.
As an example, we have calculated the optimal parameters of the driving field to generate $\pi$ and $\pi/2$ rotations via Rabi oscillations.
With this we have demonstrated coherent control of band inversion in graphene nanoribbons as well as the preparation of superposition states where strong alternating photocurrents are generated. 

While we work with the specific nanoribbon design, shown in Fig.~\ref{fig 1}, we emphasize that a similar approach can be performed for a wide range of nanoribbon sizes, and electrode configurations.
Moreover, an approach without the proposed electrostatic potential pattern is also feasible, by using graphene nanoribbons with rhomboidal pattern and superzigzag edges, which can be synthesized  with atomic precision~\cite{Beyer2019}.
Different variants of graphene nanoribbons with different patterns can also be synthesized~\cite{Chen2020, Wang2021, Alcon2025}, to which our driving field optimization methodology can also be applied.

Our proposal contributes to the development of novel coherent electronics and information processing tasks using solid-state devices.
The methodology of flattening the lowest-energy bands and finding the optimal driving field is directly relevant for designing Ramsey interferometers, spin-echo realizations, and quantum memories using graphene nanoribbons. 

\section{Acknowledgments}
We acknowledge funding by the Deutsche Forschungsgemeinschaft (DFG, German Research Foundation) “SFB-925” Project No. 170620586 and the Cluster of Excellence “Advanced Imaging of Matter” (EXC 2056), Project No. 390715994. The project is co-financed by ERDF of the European Union and by ’Fonds of the Hamburg Ministry of Science, Research, Equalities and Districts (BWFGB)’.

\bibliography{literatur}

\begin{appendix}
\section{Dependence of the band gap on the gate voltage}\label{appendix A}
\begin{figure}[h]
  \includegraphics[width=0.48\textwidth]{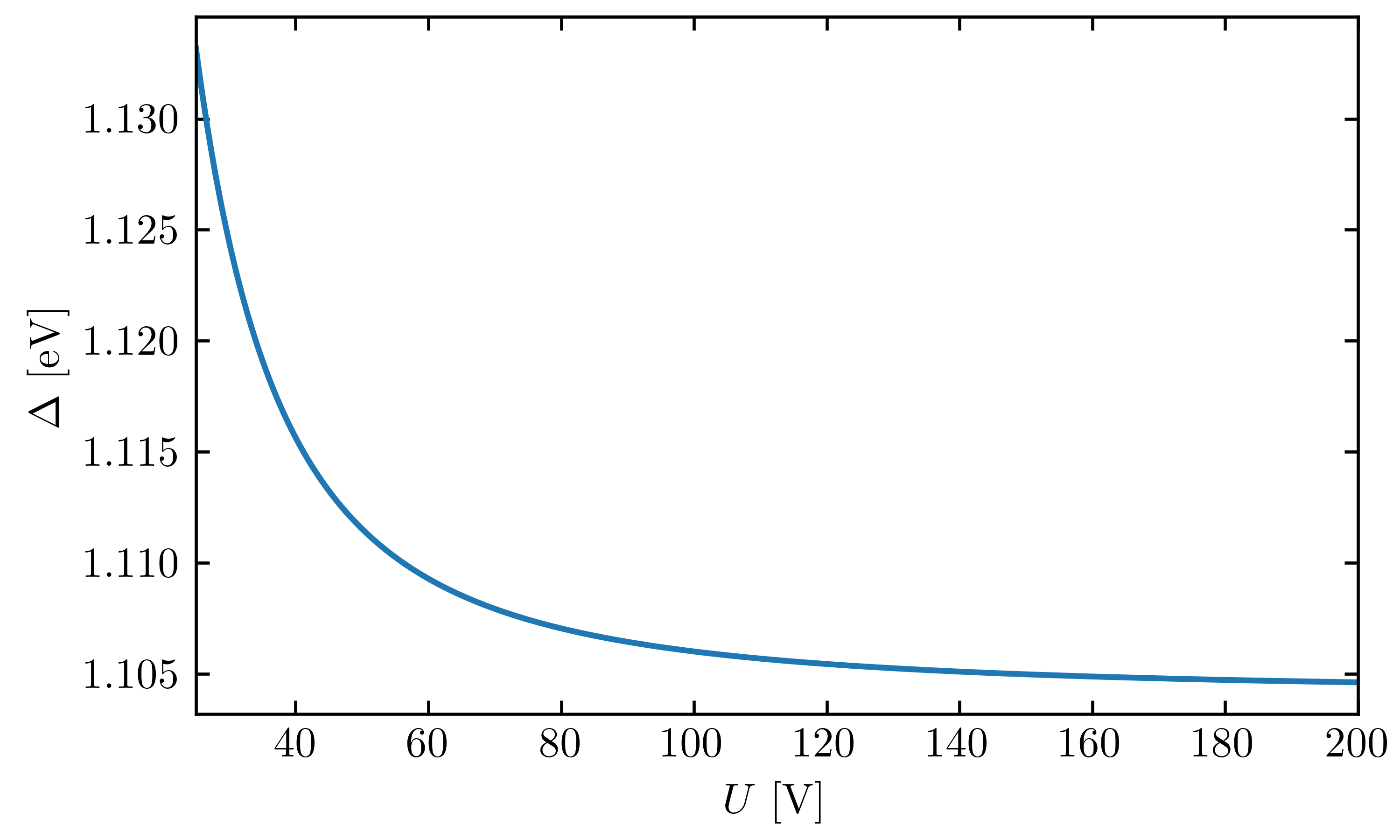}
  \caption{Dependence of the band gap on the gate voltage. The band gap is calculated by $\Delta = 2 H_{k=0,z}(\boldsymbol{A}=0)$. For increasing gate voltage the band gap is converging to $\Delta = 1.104\ \si{eV}$.}
  \label{fig 8}
\end{figure}

The dependence of the band gap on the gate voltage is shown in Fig.~\ref{fig 8}, where we calculate the band gap numerically using $\Delta = 2 H_{k=0,z}(\boldsymbol{A}=0)$.

\section{Effective $2\times 2$ Hamiltonian} \label{appendix B}
We show here the form of the effective $2\times 2$ Hamiltonian Eq.~(\ref{eqn 11}).
We take advantage of the symmetries of the system, which translate into the following equations for the matrices $\widetilde{h}^{(\lambda)^{\mathrm{eff}}}$,
\begin{eqnarray}
  \text{Im}\ \widetilde{h}_{00}^{(0)^{\mathrm{eff}}} &=& \text{Im}\ \widetilde{h}_{11}^{(0)^{\mathrm{eff}}} = 0 \label{eqn B1}\\
  \widetilde{h}_{00}^{(1)^{\mathrm{eff}}} &=&  \overline{\widetilde{h}_{00}^{(2)^{\mathrm{eff}}}}\\
  \widetilde{h}_{11}^{(1)^{\mathrm{eff}}} &=&  \overline{\widetilde{h}_{11}^{(2)^{\mathrm{eff}}}}\\
  \widetilde{h}_{00}^{(3)^{\mathrm{eff}}} = \overline{\widetilde{h}_{00}^{(4)^{\mathrm{eff}}}} &=& \widetilde{h}_{00}^{(5)^{\mathrm{eff}}} = \overline{\widetilde{h}_{00}^{(6)^{\mathrm{eff}}}}\\
  \widetilde{h}_{11}^{(3)^{\mathrm{eff}}} = \overline{\widetilde{h}_{11}^{(4)^{\mathrm{eff}}}} &=& \widetilde{h}_{11}^{(5)^{\mathrm{eff}}} = \overline{\widetilde{h}_{11}^{(6)^{\mathrm{eff}}}}\\
    \widetilde{h}_{01}^{(0)^{\mathrm{eff}}} &\approx& 0. \label{eqn B6}
\end{eqnarray}
Here the overlines denote the complexe conjugate.
We rewrite the effective Hamiltonian Eq.~(\ref{eqn 11}) using Eq.~(\ref{eqn 6}) and Eq.~(\ref{eqn B1})~-~(\ref{eqn B6}) to obtain
\begin{widetext}
\begin{eqnarray}
  H_{k}^{\mathrm{eff}} &=& \left[\frac{1}{2}U\left(\widetilde{h}_{00}^{(0)^{\mathrm{eff}}} + \widetilde{h}_{11}^{(0)^{\mathrm{eff}}}\right) -|t|\left(\frac{1}{2}\left(\widetilde{h}_{00}^{(1)^{\mathrm{eff}}} + \widetilde{h}_{11}^{(1)^{\mathrm{eff}}}\right) e^{2i\phi_x}
      + \frac{1}{2}\left(\overline{\widetilde{h}_{00}^{(1)^{\mathrm{eff}}}} + \overline{\widetilde{h}_{11}^{(1)^{\mathrm{eff}}}}\right) e^{-2i\phi_x} \right.\right.\nonumber\\
  &&+\left.\left. \frac{1}{2}\left(\widetilde{h}_{00}^{(3)^{\mathrm{eff}}} + \widetilde{h}_{11}^{(3)^{\mathrm{eff}}}\right) \left(e^{i(\phi_x + \phi_y)} + e^{i(\phi_x - \phi_y)}\right)
      + \frac{1}{2}\left(\overline{\widetilde{h}_{00}^{(3)^{\mathrm{eff}}}} + \overline{\widetilde{h}_{11}^{(3)^{\mathrm{eff}}}}\right) \left(e^{-i(\phi_x + \phi_y)} + e^{-i(\phi_x - \phi_y)}\right)\right)\right] \cdot 1\kern-0.25em\text{l} \nonumber\\
  &&-\ |t|\ \text{Re}\ \left(\widetilde{h}_{01}^{(1)^{\mathrm{eff}}} e^{2i\phi_x} + \widetilde{h}_{01}^{(2)^{\mathrm{eff}}} e^{-2i\phi_x}
      + \widetilde{h}_{01}^{(3)^{\mathrm{eff}}} e^{i(\phi_x + \phi_y)} + \widetilde{h}_{01}^{(4)^{\mathrm{eff}}} e^{-i(\phi_x + \phi_y)}
      + \widetilde{h}_{01}^{(5)^{\mathrm{eff}}} e^{i(\phi_x - \phi_y)} + \widetilde{h}_{01}^{(6)^{\mathrm{eff}}} e^{-i(\phi_x - \phi_y)} \right) \sigma_x \nonumber\\
  &&+\ |t|\ \text{Im}\ \left(\widetilde{h}_{01}^{(1)^{\mathrm{eff}}} e^{2i\phi_x} + \widetilde{h}_{01}^{(2)^{\mathrm{eff}}} e^{-2i\phi_x}
      + \widetilde{h}_{01}^{(3)^{\mathrm{eff}}} e^{i(\phi_x + \phi_y)} + \widetilde{h}_{01}^{(4)^{\mathrm{eff}}} e^{-i(\phi_x + \phi_y)}
      + \widetilde{h}_{01}^{(5)^{\mathrm{eff}}} e^{i(\phi_x - \phi_y)} + \widetilde{h}_{01}^{(6)^{\mathrm{eff}}} e^{-i(\phi_x - \phi_y)} \right) \sigma_y \nonumber\\
  &&+\left[\frac{1}{2}U\left(\widetilde{h}_{00}^{(0)^{\mathrm{eff}}} - \widetilde{h}_{11}^{(0)^{\mathrm{eff}}}\right) - |t|\left(\frac{1}{2}\left(\widetilde{h}_{00}^{(1)^{\mathrm{eff}}} - \widetilde{h}_{11}^{(1)^{\mathrm{eff}}}\right) e^{2i\phi_x}
      + \frac{1}{2}\left(\overline{\widetilde{h}_{00}^{(1)^{\mathrm{eff}}}} - \overline{\widetilde{h}_{11}^{(1)^{\mathrm{eff}}}}\right) e^{-2i\phi_x}\right.\right.\nonumber\\
  &&+\left.\left. \frac{1}{2}\left(\widetilde{h}_{00}^{(3)^{\mathrm{eff}}} - \widetilde{h}_{11}^{(3)^{\mathrm{eff}}}\right) \left(e^{i(\phi_x + \phi_y)} + e^{i(\phi_x - \phi_y)}\right)
      + \frac{1}{2}\left(\overline{\widetilde{h}_{00}^{(3)^{\mathrm{eff}}}} - \overline{\widetilde{h}_{11}^{(3)^{\mathrm{eff}}}}\right) \left(e^{-i(\phi_x + \phi_y)} + e^{-i(\phi_x - \phi_y)}\right)\right)\right] \cdot \sigma_z. \nonumber\\
\end{eqnarray}
\end{widetext}

\section{Calculation of the linear polarization angle}\label{appendix C}
In the following we calculate, for the momentum $k=0$, the optimal linear polarization angle for a Rabi-like driving pulse.
For $k=0$ all coefficient matrices $\widetilde{h}^{(\lambda)^{\mathrm{eff}}}$ are real and Eq.~(\ref{eqn B1})~-~(\ref{eqn B6}) simplify further to
\begin{eqnarray}
  \widetilde{h}_{00}^{(1)^{\mathrm{eff}}} = \widetilde{h}_{00}^{(2)^{\mathrm{eff}}} \label{eqn C1}\\
  \widetilde{h}_{00}^{(3)^{\mathrm{eff}}} = \widetilde{h}_{00}^{(4)^{\mathrm{eff}}} = \widetilde{h}_{00}^{(5)^{\mathrm{eff}}} = \widetilde{h}_{00}^{(6)^{\mathrm{eff}}} \label{eqn C2}
\end{eqnarray}
and
\begin{eqnarray}
  \widetilde{h}_{01}^{(0)^{\mathrm{eff}}} = 0 \label{eqn C3}\\
  \widetilde{h}_{01}^{(1)^{\mathrm{eff}}} = -\widetilde{h}_{01}^{(2)^{\mathrm{eff}}} \label{eqn C4}\\
  \widetilde{h}_{01}^{(3)^{\mathrm{eff}}} = -\widetilde{h}_{01}^{(4)^{\mathrm{eff}}} = \widetilde{h}_{01}^{(5)^{\mathrm{eff}}} = -\widetilde{h}_{01}^{(6)^{\mathrm{eff}}} \label{eqn C5}\\
  \widetilde{h}_{10}^{(\lambda)^{\mathrm{eff}}} = -\widetilde{h}_{01}^{(\lambda)^{\mathrm{eff}}}
\end{eqnarray}

We find the optimal linear polarization angle $\vartheta$ by the condition
\begin{eqnarray}
  H_{k,z}(A_x, A_y) = H_{k,z}(0, 0),
\end{eqnarray}
which we can rewrite equivalently as
\begin{widetext}
\begin{eqnarray}
  \frac{1}{2} \left(H_{k,00}^{\mathrm{eff}}(A_x, A_y) - H_{k,11}^{\mathrm{eff}}(A_x, A_y)\right) = \frac{1}{2} \left(H_{k,00}^{\mathrm{eff}}(0,0) - H_{k,11}^{\mathrm{eff}}(0,0)\right). \label{eqn C8}
\end{eqnarray}
For $H_{k,00}^{\mathrm{eff}}(A_x, A_y)$ we use the decomposition Eq.~(\ref{eqn 6}) together with Eq.~(\ref{eqn C1}) and Eq.~(\ref{eqn C2}) to get
\begin{eqnarray}
  H_{k,00}^{\mathrm{eff}}(A_x, A_y) &=& U \widetilde{h}_{00}^{(0)^{\mathrm{eff}}} - |t|\left[\widetilde{h}_{00}^{(1)^{\mathrm{eff}}} \left(e^{2i\phi_x} + e^{-2i\phi_x}\right) + \widetilde{h}_{00}^{(3)^{\mathrm{eff}}} \left(e^{i(\phi_x + \phi_y)} + e^{-i(\phi_x + \phi_y)} + e^{i(\phi_x - \phi_y)} + e^{-i(\phi_x - \phi_y)}\right) \right] \nonumber\\
  &=& U \widetilde{h}_{00}^{(0)^{\mathrm{eff}}} - |t|\left[2 \widetilde{h}_{00}^{(1)^{\mathrm{eff}}} \cos(2\phi_x) + 4 \widetilde{h}_{00}^{(3)^{\mathrm{eff}}} \cos(\phi_x) \cos(\phi_y)\right].
\end{eqnarray}
And for $H_{k,11}^{\mathrm{eff}}(A_x, A_y)$ we get in the same way 
\begin{eqnarray}
  H_{k,11}^{\mathrm{eff}}(A_x, A_y) = U \widetilde{h}_{11}^{(0)^{\mathrm{eff}}} - |t|\left[2 \widetilde{h}_{11}^{(1)^{\mathrm{eff}}} \cos(2\phi_x) + 4 \widetilde{h}_{11}^{(3)^{\mathrm{eff}}} \cos(\phi_x) \cos(\phi_y)\right].
\end{eqnarray}
Therefore, the condition Eq.~(\ref{eqn C8}) transforms to
\begin{eqnarray}
  &&U \left(\widetilde{h}_{00}^{(0)^{\mathrm{eff}}} - \widetilde{h}_{11}^{(0)^{\mathrm{eff}}}\right) - |t|\left[2 \left(\widetilde{h}_{00}^{(1)^{\mathrm{eff}}} - \widetilde{h}_{11}^{(1)^{\mathrm{eff}}}\right) \cos(2\phi_x) + 4 \left(\widetilde{h}_{00}^{(3)^{\mathrm{eff}}} - \widetilde{h}_{11}^{(3)^{\mathrm{eff}}}\right) \cos(\phi_x) \cos(\phi_y)\right] \nonumber\\
  &=& U \left(\widetilde{h}_{00}^{(0)^{\mathrm{eff}}} - \widetilde{h}_{11}^{(0)^{\mathrm{eff}}}\right) - |t|\left[2 \left(\widetilde{h}_{00}^{(1)^{\mathrm{eff}}} - \widetilde{h}_{11}^{(1)^{\mathrm{eff}}}\right) + 4 \left(\widetilde{h}_{00}^{(3)^{\mathrm{eff}}} - \widetilde{h}_{11}^{(3)^{\mathrm{eff}}}\right)\right],
\end{eqnarray}
\end{widetext}
which can be solved for $\phi_y$ as
\begin{eqnarray}
  \phi_y = \arccos\left(\frac{\widetilde{h}_{00}^{(1)^{\mathrm{eff}}} - \widetilde{h}_{11}^{(1)^{\mathrm{eff}}}}{2\left(\widetilde{h}_{00}^{(3)^{\mathrm{eff}}} - \widetilde{h}_{11}^{(3)^{\mathrm{eff}}}\right)} \frac{1 - \cos(2\phi_x)}{\cos(\phi_x)} + \frac{1}{\cos(\phi_x)}\right). \nonumber\\
\end{eqnarray}
We approximate this solution for small vector potentials by
\begin{eqnarray}
  \phi_y = \sqrt{-\left(1 + 2\ \frac{\widetilde{h}_{00}^{(1)^{\mathrm{eff}}} - \widetilde{h}_{11}^{(1)^{\mathrm{eff}}}}{\widetilde{h}_{00}^{(3)^{\mathrm{eff}}} - \widetilde{h}_{11}^{(3)^{\mathrm{eff}}}}\right)} \phi_x + O[\phi_x^3].
\end{eqnarray}
By defining $\gamma = \sqrt{-\left(1 + 2\ \frac{\widetilde{h}_{00}^{(1)^{\mathrm{eff}}} - \widetilde{h}_{11}^{(1)^{\mathrm{eff}}}}{\widetilde{h}_{00}^{(3)^{\mathrm{eff}}} - \widetilde{h}_{11}^{(3)^{\mathrm{eff}}}}\right)}$ this simplifies to
\begin{eqnarray}
  \phi_y \approx \gamma \phi_x,
\end{eqnarray}
from which the linear polarization angle $\vartheta$ follows as
\begin{eqnarray}
  \vartheta = \arctan\left(\frac{\gamma}{\sqrt{3}}\right).
\end{eqnarray}

\section{Calculation of the amplitude factor}\label{appendix D}
In order to find the correct factor $A_0$ for the driving field, we have to ensure the condition
\begin{equation} \label{eqn D1}
  H_{k,01}^{\mathrm{eff}} = -i\hbar \Omega(t) \sin(\omega_{dr}t).
\end{equation}
We rewrite $H_{k,01}^{\mathrm{eff}}(A_x, A_y)$ by using Eq.~(\ref{eqn 6}) together with Eq.~(\ref{eqn C3}), Eq.~(\ref{eqn C4}) and Eq.~(\ref{eqn C5}) as
\begin{widetext}
\begin{eqnarray}
  H_{k,01}^{\mathrm{eff}}(A_x, A_y) &=& - |t| \left[\widetilde{h}_{01}^{(1)^{\mathrm{eff}}} \left(e^{2i\phi_x} - e^{-2i\phi_x}\right) + \widetilde{h}_{01}^{(3)^{\mathrm{eff}}} \left(e^{i(\phi_x + \phi_y)} - e^{-i(\phi_x + \phi_y)} + e^{i(\phi_x - \phi_y)} - e^{-i(\phi_x - \phi_y)}\right)\right] \\
  &=& - |t| \left[\widetilde{h}_{01}^{(1)^{\mathrm{eff}}}\ 2i\ \sin(2\phi_x) + \widetilde{h}_{01}^{(3)^{\mathrm{eff}}}\ 4i\ \sin(\phi_x)\cos(\phi_y)\right] \\
  &\approx& - |t| \left[\widetilde{h}_{01}^{(1)^{\mathrm{eff}}}\ 4i\ \phi_x + \widetilde{h}_{01}^{(3)^{\mathrm{eff}}}\ 4i\ \phi_x\right] \\
  &=& -4i|t| \left(\widetilde{h}_{01}^{(1)^{\mathrm{eff}}} + \widetilde{h}_{01}^{(3)^{\mathrm{eff}}}\right) \phi_x,
\end{eqnarray}
where the approximation is valid for small vector potentials $|\boldsymbol{A}| \ll \frac{2\hbar}{ea_0}$.
We replace $\phi_x$ by using Eq.~(\ref{eqn 7}) and Eq.~(\ref{eqn 13}) and obtain
\begin{eqnarray}
  H_{k,01}^{\mathrm{eff}} &=& -4i|t| \left(\widetilde{h}_{01}^{(1)^{\mathrm{eff}}} + \widetilde{h}_{01}^{(3)^{\mathrm{eff}}}\right) \left(-\frac{ea_0}{2\hbar}\right) A_0 \cos(\vartheta) \hbar \Omega(t) 2\sin(\omega_{dr}t) \\
  &=& i\frac{2ea_0|t|}{\hbar} \left(\widetilde{h}_{01}^{(1)^{\mathrm{eff}}} + \widetilde{h}_{01}^{(3)^{\mathrm{eff}}}\right) A_0 \cos(\vartheta) \hbar \Omega(t) 2\sin(\omega_{dr}t).
\end{eqnarray}
Thereby, the condition Eq.~(\ref{eqn D1}) transforms to
\begin{eqnarray}
  i\frac{2ea_0|t|}{\hbar} \left(\widetilde{h}_{01}^{(1)^{\mathrm{eff}}} + \widetilde{h}_{01}^{(3)^{\mathrm{eff}}}\right) A_0 \cos(\vartheta) \hbar \Omega(t) 2\sin(\omega_{dr}t) = -i\hbar \Omega(t) \sin(\omega_{dr}t).
\end{eqnarray}
\end{widetext}
We solve the condition for the factor $A_0$ by
\begin{eqnarray}
  A_0 = -\frac{\hbar}{4ea_0|t|}\frac{1}{\cos(\vartheta)}\frac{1}{\widetilde{h}_{01}^{(1)^{\mathrm{eff}}} + \widetilde{h}_{01}^{(3)^{\mathrm{eff}}}}.
\end{eqnarray}

\end{appendix}

\end{document}